\documentclass[twocolumn,showpacs,preprintnumbers,amsmath,amssymb,aps]{revtex4-1}
\usepackage{graphicx}
\usepackage{dcolumn}
\usepackage{bm}
\usepackage{float}
\usepackage{soul}

\begin{document}

\title{Improved thermometry of low-temperature quantum systems by a ring-structure probe}

\author{Li-Sha Guo}
\author{Bao-Ming Xu}
\author{Jian Zou}%
\email{zoujian@bit.edu.cn}
\author{Bin Shao}%

\affiliation{School of Physics, Beijing Institute of Technology, Beijing 100081, China}%

\date{Submitted \today}

\begin{abstract}
The thermometry precision of a sample is a question of both fundamental and technological importance. In this paper, we consider a ring-structure system as our probe to estimate the temperature of a bath. Based on the Markovian master equation of the probe, we calculate the quantum Fisher information (QFI) of the probe at any time. We find that for the thermal equilibrium thermometry, the ferromagnetic structure can measure a lower temperature of the bath with a higher precision compared with the non-structure probe. While for the dynamical thermometry, the antiferromagnetic structure can make the QFI of the probe in the dynamical process much larger than that in equilibrium with the bath, which is somewhat counterintuitive. Moreover, the best accuracy for the thermometry achieved in the antiferromagnetic structure case can be much higher than that in the non-structure case. The physical mechanisms of above phenomena are given in this paper.
\end{abstract}


\maketitle

\section{Introduction}
Parameter estimation is a fundamental and important subject in physics, with its applications in various aspects, such as gravitational-wave detectors \cite{Caves1981,McKenzie2002}, frequency spectroscopy \cite{Wineland1992,Bollinger1996}, interferometry \cite{Holland1993,Lee2002}, and atomic clocks \cite{Valencia2004,de Burgh2005}. And one usually utilizes Cram\'{e}r-Rao bound \cite{Cramer1999} on the error as a criterion to assess the performance of a parameter estimation technique, which is proportional to the inverse of the square root of the so-called Fisher information (FI) \cite{Cramer1999,Fisher1922,Rao1973}. The maximization of the FI over all measurement strategies allowed by quantum mechanics lead to a nontrivial quantity: quantum Fisher information (QFI).

Temperature, being one of the most fundamental and the most frequently measured physical quantity, has recently attracted a growing interest in obtaining an accurate reading. Indeed, precise knowledge of the temperature of a sample proved indispensable for many advancements in physics \cite{Ruostekoski2009}, biology \cite{Kucsko2013}, material science \cite{Toyli2013} and microelectronic industry \cite{Jiang2003}. Also the task of temperature measurement can be translated using the language of estimation theory to the problem of parameter estimation.

With the progress in manipulation of individual quantum system, the study of thermometry precision, using individual quantum system as a probe, has attracted considerable attentions \cite{Correa2015,Brunelli2011,Brunelli2012,Jevtic2015,Higgins2013,Bruderer2006,Sabin2014}. Specifically, Ref. \cite{Correa2015} analyzed the thermometry of an unknown bath and proved that the optimal quantum probe is an effective two-level atom with a maximally degenerate excited state, while Refs. \cite{Brunelli2011,Brunelli2012} used a single qubit as the probe to estimate the temperature of the micro-mechanical resonators. Meanwhile, Jevtic et al. \cite{Jevtic2015} have also used a single qubit to distinguish between two different temperatures of a bosonic bath and found the potential role played by coherence and entanglement in simple thermometric tasks. In addition, Ref. \cite{Higgins2013} has made use of the ac Stark effect to implement the practical and precise qubit thermometry of an oscillator. And Refs. \cite{Bruderer2006,Sabin2014} used two-level atomic quantum dots as thermometers of BECs. The fundamental and important questions such as the scaling of the precision of temperature estimation with the number of quantum probes has been discussed in Ref. \cite{Stace2010}, and it was shown that it is possible to map the problem of measuring the temperature onto the problem of estimating an unknown phase, as a result, the scaling of the precision of a thermometer may be in principle improved to ¡«$1/N$, representing the Heisenberg limit to thermometry. Following this paper, Jarzyna and Zwierz provided a detailed description of the interferometric thermometer and found that this approach is capable of measuring the temperature of a sample in the nK regime \cite{Jarzyna2014}. Recently, Ref. \cite{Pasquale2015} introduced the local quantum thermal susceptibility (LQTS) functional to quantify the best achievable accuracy for the temperature estimation of a composite system via local measurements. And Ref. \cite{Kliesch2014} has clarified the limitations of a universal concept of scale-independent temperature by showing that temperature is intensive on a given length scale if and only if correlations are negligible. Besides, some theoretical works \cite{Boixo2007,Choi2008,Roy2008} have shown that interactions among particles may be a valuable resource for quantum metrology, allowing scaling beyond the Heisenberg limit. Recently, Ref. \cite{Mehboudi2015} has used an ultracold lattice gas simulating a strongly correlated system consisted of $N$ interacting particles as a probe for the thermometry.

In this paper, we consider a ring-structure system with nearest neighbor interactions as our probe to estimate the temperature of an electromagnetic field (bath). And the ring-structure probe, consisted of $N$ two-level atoms, is coupled through dipole interactions with the electromagnetic field. We first calculate the QFI of the probe at any time and then analyze how the structure (strength of the dipole-dipole interaction between adjacent atoms) and the particle number of the probe affect the temperature estimation in two complementary scenarios, i.e., the thermal equilibrium thermometry (one estimates the temperature when the probe reaches thermal equilibrium with the bath) and the dynamical thermometry (one estimates the temperature before the probe  attains full thermalization). For the dynamical thermometry, we use the Greenberger-Horne-Zeilinger (GHZ) state as the initial state of our probe and study its dynamical evolution before achieving full thermalization with the electromagnetic field.

Our main results are the following. First, for the thermal equilibrium thermometry, the ferromagnetic probe can measure a lower temperature of the bath with an improved precision compared with the non-structure case. More accurately, when the structure is ferromagnetic, as the absolute value of the coupling strength increases, the optimal temperature $T_{opt}$, at which the QFI achieves its maximum, becomes lower and the value of the corresponding equilibrated QFI of the probe becomes larger. However, the probe would take a longer time to be equilibrated with the bath. Fortunately, we can reduce this time by increasing the particle number $N$ of the probe. In contrast, for the dynamical thermometry, the antiferromagnetic structure would play a distinctive role. Specifically, when the coupling strength increases to a certain value, the QFI of the probe in the dynamical process can be larger than that in equilibrium with the bath, which is somewhat counterintuitive. Besides, the best accuracy for the thermometry achieved in the antiferromagnetic structure case can be much higher than that in the non-structure case, moreover, the larger the coupling strength, the lower the $T_{opt}$ and the larger the optimal QFI, but the optimal measurement time $t_{opt}$ becomes longer. Similarly, we can reduce this optimal measurement time $t_{opt}$ by increasing the particle number $N$ of the probe.

The remainder of this paper is organized as follows: in Sec. \uppercase\expandafter{\romannumeral2} we first introduce our model and its dynamics and then we analyze its energy level structure. In Sec. \uppercase\expandafter{\romannumeral3}, we first review the quantum parameter estimation theory and then calculate the QFI at any time for our model, and based on which we analyze the effects of the structure and the particle number $N$ of the probe on the thermometry precision in two complementary scenarios, i.e., the thermal equilibrium thermometry and the dynamical thermometry. Finally Sec. \uppercase\expandafter{\romannumeral4} closes the paper with some concluding remarks.

\begin{figure}[tbp]
\begin{center}
\includegraphics[width=9cm]{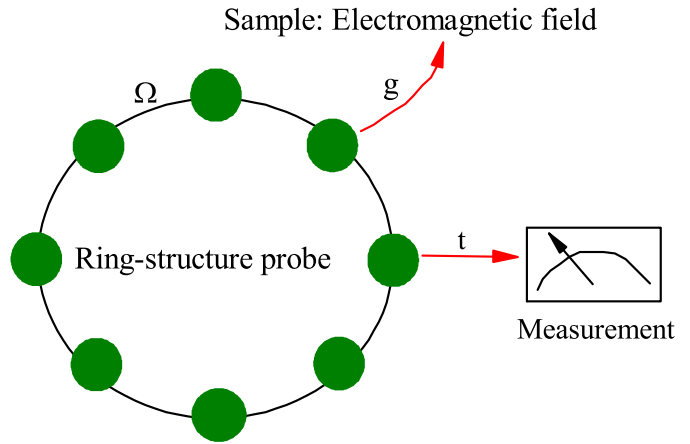}
\caption {(Color online) A schematic diagram showing the thermometry process with a ring-structure probe. $\Omega$ represents the coupling between adjacent atoms of the probe, $g$ denotes the coupling between the probe and the electromagnetic field, and $t$ represents the evolution time of the probe at which the probe is measured to estimate the temperature $T$ of the electromagnetic field.}
\end{center}
\end{figure}

\section{Ring-structure probe and its dynamical evolution}
\subsection{Model and Dynamics}
In this paper, we adopt a ring structure \cite{Gross1982,Higgins2014} system consisted of $N$ identical and permutational symmetric two-level atoms (see Fig. 1) as a probe to detect the temperature of an electromagnetic field. The Hamiltonian of $N$ atoms is ($\hbar=c=1$)
\begin{equation}\label{HS}
    H_{s}= \frac{1}{2}\omega_{A}\sum_{n=1}^{N}\sigma_{z}^{n}
\end{equation}
with $\omega_{A}$ being the bare atomic transition frequency and $\sigma_{z}^{n}=|e\rangle\langle e|-|g\rangle\langle g|$ being the Pauli operator for the $n$th atom. The Hamiltonian of the electromagnetic field is
\begin{equation}\label{HB}
    H_{B}=\sum_{\textbf{k}}\sum_{\lambda=1,2}\omega_{\textbf{k}}a_{\lambda}^{\dag}(\textbf{k})a_{\lambda}(\textbf{k}),
\end{equation}
where $\omega_{\textbf{k}}$ is the field frequency for the wave vector $\textbf{k}$, and $a_{\lambda}(\textbf{k})$ and $a_{\lambda}^{\dag}(\textbf{k})$ are the field annihilation and creation operators, respectively. $\lambda=1,2$ denote two independent polarization directions of the electromagnetic field for each $\textbf{k}$. The atom-field interaction can be expressed as
\begin{equation}\label{HI}
    H_{I}=-\sum^{N}_{n}\bigl(\sigma_{-}^{n}\textbf{d}\cdot \hat{\textbf{E}}(\textbf{r}_{n})+\sigma_{+}^{n}\textbf{d}^{\ast}\cdot \hat{\textbf{E}}(\textbf{r}_{n})\bigr),
\end{equation}
in which $\sigma_{+}^{n}=|e\rangle\langle g|$ and $\sigma_{-}^{n}=|g\rangle\langle e|$ are the upper and lower operators to describe the $n$th two-level atom, and $\mathbf{d}$ is the atomic dipole vector. The electric field operator $\hat{\textbf{E}}$ is given by
\begin{equation}
  \hat{\textbf{E}}(\textbf{r}_{n})=i\sum_{n=1}^{N}\sum_{\textbf{k},\lambda}\sqrt{\frac{2\pi\omega_{k}}{V}}\textbf{e}_{\lambda}(\textbf{k})
  \bigl(a_{\lambda}^{\dag}(\textbf{k})e^{-i\textbf{k}\cdot\textbf{r}_{n}}-a_{\lambda}(\textbf{k})e^{i\textbf{k}\cdot\textbf{r}_{n}}\bigr),
\end{equation}
where, $\textbf{e}_{\lambda}(\textbf{k})$ is the polarization vector of the field, $V$ is an arbitrary quantization volume, much larger than the atomic system and $\textbf{r}_{n}$ is the position vector of the $n$th two-level atom.

The system dynamics is then generically determined by the following master equation (the detailed derivation is shown in Appendix A) \cite{Gross1982,Higgins2014,Breuer2002}£º
\begin{widetext}
\begin{equation}\label{EQ}
\begin{split}
\frac{d\rho_{s}(t)}{dt}=&-i[H_{s}+H_{d},\rho_{s}(t)]\\
&+\sum_{\omega>0}\sum^{N}_{m,n}\frac{4\omega^{3}|d|^2}{3}
\biggl(\bigl(1+N(\omega)\bigr)\bigl(\sigma^{n}_{-}\rho_{s}(t)\sigma^{m}_{+}-\frac{1}{2}\big\{\sigma^{m}_{+}\sigma^{n}_{-},\rho_{s}(t)\big\}\bigr)
+N(\omega)\bigl(\sigma^{n}_{+}\rho_{s}(t)\sigma^{m}_{-}-\frac{1}{2}\big\{\sigma^{m}_{-}\sigma^{n}_{+},\rho_{s}(t)\big\}\bigr)\biggr),
\end{split}
\end{equation}
\end{widetext}
where $N(\omega)=(\exp[\omega/T]-1)^{-1}$ is the Planck distribution with $T$ being the temperature, and satisfies $N(-\omega)=-(1+N(\omega))$ (here we let the Boltzmann constant $k_{B}=1$).
The Hamiltonian
\begin{equation}\label{Hd}
  H_{d}=\Omega\sum_{n}(\sigma^{n}_{+}\sigma^{n+1}_{-}+\sigma^{n}_{-}\sigma^{n+1}_{+})
\end{equation}
is the Van der Waals dipole-dipole interaction induced by the electromagnetic field where $\Omega$ is the interaction strength (throughout this paper, the term `structure' refers to it), and the periodic boundary condition $\sigma^{N+1}_{\pm}=\sigma^{1}_{\pm}$ is considered. Due to the fact that $[H_{s},H_{d}]=0$, so to the first order of the Van der Waals dipole-dipole interaction, i.e., Eq. (\ref{Hd}), it does not mix the eigenstates of $H_{s}$, only shifts their energies. And it is this energy level shift that could contribute to the thermometry precision, which will be showed in the next section. While in the following, we will discuss this energy shift in detail.

\subsection{Energy Levels}
Due to the permutation symmetry of the probe system, the $N$ atoms become indistinguishable such that the electromagnetic field interacts with them collectively. The dynamics is then best described
by collective operators:
\begin{equation}\label{J+}
    J_{\pm}=\sum_{n=1}^{N}\sigma_{\pm}^{n},~~~~~  J_{z}=\frac{1}{2}\sum_{n=1}^{N}\sigma_{z}^{n}.
\end{equation}
And any $N$ spin-1/2 state invariant by atom permutation is an eigenstate corresponding to the maximum $J=N/2$ value of the angular momentum, which can be written as the Dicke state $|J,M\rangle$, obtained by repeated action of the symmetrical collective deexcitation operator $J_{-}$ on the state $|e,e\cdots e\rangle$:
\begin{equation}\label{Dicke state}
    |J,M\rangle=\sqrt{\frac{(J+M)!}{N!(J-M)!}}J_{-}^{J-M}|e,e\cdots e\rangle.
\end{equation}
And the actions of the collective operators $J_{\pm}$ and $J_{z}$ on the Dicke state can be described as:
\begin{equation}\label{action}
    J_{\pm}|J,M\rangle=\sqrt{(J\pm M+1)(J\mp M)}|J,M\pm1\rangle,
\end{equation}
and
\begin{equation}\label{jz}
    J_{z}|J,M\rangle=M|J,M\rangle.
\end{equation}
For convenience, we would write $|M\rangle$ instead of $|J,M\rangle$ for $J=N/2$ in the following.

Now let us analyze how the energy level would change in the presence of the Van der Waals dipole-dipole interaction (Eq. (\ref{Hd})) between the adjacent atoms of the probe. Due to the fact that $H_{d}$ (Eq. (\ref{Hd})) does not mix the eigenstates of $H_{s}$, only shifts their energies as mentioned above, the effective Hamiltonian of the probe can be written as a diagonalized form \cite{Gross1982,Higgins2014}:
\begin{equation}
  H_{e}\equiv H_{s}+H_{d}=\sum_{M=-J}^{J}E_{M}|M\rangle\langle M|,
\end{equation}
where $E_{M}$ is the eigenenergy of the state $|M\rangle$:
\begin{equation}\label{EM}
    E_{M}=M\omega_{A}+\Omega\frac{J^{2}-M^{2}}{J-\frac{1}{2}}.
\end{equation}
And the energy difference between any two adjacent energy levels can be obtained as:
\begin{equation}\label{wM}
\begin{split}
    \Delta E_{M\rightarrow M-1}&\equiv E_{M}-E_{M-1}\\
    &=\omega_{A}-4\Omega\frac{M-\frac{1}{2}}{N-1}.
\end{split}
\end{equation}
From Eq. (\ref{wM}) we can see that if we neglect the dipole-dipole interaction, i.e., $\Omega=0$, the transition frequencies between any two adjacent energy levels are degenerate and equal to $\omega_{A}$. On the contrary, if we consider the dipole-dipole interaction, i.e., $\Omega\neq0$, then it would break the degeneracy of the transition frequency.

\begin{center}
\includegraphics[width=8cm]{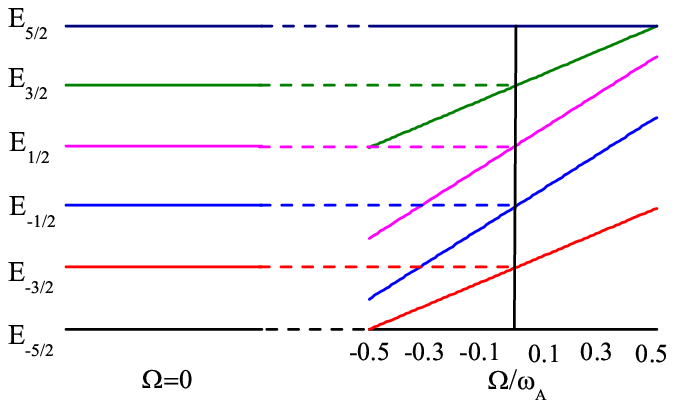}
\parbox{8cm}{\small{FIG. 2.} (Color online) Schematic drawing of the Hamiltonian spectrum (Eq. (\ref{EM})) as a function of the coupling strength $\Omega$ for particle number $N$=5 as an example. Here we take $\omega_{A}$ as the unit, i.e., $\omega_{A}=1$.}
\end{center}

Fig. 2 is a paradigm schematic drawing of the Hamiltonian spectrum (Eq. (\ref{EM})) as a function of the coupling strength $\Omega$ with $N=5$. We can see that when there is no interactions ($\Omega=0$), the energy level is equally spaced, i.e., each transition has the same frequency $\omega_{A}$ (Throughout this paper, we take $\omega_{A}$ as unit 1). But as the coupling strength $\Omega$ changes, each energy level would shift except for the ground level and the highest excited level, and become unequally spaced, i.e., each transition now has a unique frequency $\omega_{M}=\Delta E_{M\rightarrow M-1}$. We emphasize that we only consider the range $\Omega\in(-0.5,~0.5)$ because the Hamiltonian spectrum (Eq. (\ref{EM})) would become very complicated and the energy levels would have at least one energy level crossing for $\Omega\geq0.5$ or $\Omega\leq-0.5$.

And then we would analyze the effects of the particle number $N$ and the coupling strength $\Omega$ on the energy difference $\Delta E_{M\rightarrow M-1}$, which will be used in Sec. \uppercase\expandafter{\romannumeral3}.
From Eq. (\ref{wM}), we can obtain that the energy difference between the two highest energy levels $E_{\frac{N}{2}}$ and $E_{\frac{N}{2}-1}$ is $\Delta E_{\frac{N}{2}\rightarrow\frac{N}{2}-1}\equiv E_{\frac{N}{2}}-E_{\frac{N}{2}-1}=\omega_{A}-2\Omega$ and the energy difference between the two lowest energy levels $E_{-\frac{N}{2}+1}$ and $E_{-\frac{N}{2}}$ is $\Delta E_{-\frac{N}{2}+1\rightarrow-\frac{N}{2}}\equiv E_{-\frac{N}{2}+1}-E_{-\frac{N}{2}}=\omega_{A}+2\Omega$, and we can see that both $\Delta E_{\frac{N}{2}\rightarrow\frac{N}{2}-1}$ and $\Delta E_{-\frac{N}{2}+1\rightarrow-\frac{N}{2}}$ are independent of the particle number $N$. Moreover, for the higher energy differences near $\Delta E_{\frac{N}{2}\rightarrow\frac{N}{2}-1}$, we can obtain that $\Delta E_{H}\sim E_{\frac{N}{2}-1}-E_{\frac{N}{2}-2}=\omega_{A}-2\Omega(1-\frac{2}{N-1})=\Delta E_{\frac{N}{2}\rightarrow\frac{N}{2}-1}+\frac{4\Omega}{N-1}$, and for the lower energy differences near $\Delta E_{-\frac{N}{2}+1\rightarrow-\frac{N}{2}}$, we can obtain that $\Delta E_{L}\sim E_{-\frac{N}{2}+2}-E_{-\frac{N}{2}+1}=\omega_{A}+2\Omega(1-\frac{2}{N-1})=\Delta E_{-\frac{N}{2}+1\rightarrow-\frac{N}{2}}-\frac{4\Omega}{N-1}$. Here, the subscripts $H$ and $L$ on $\Delta E$ mean `High' and `Low', respectively.
From the expressions of $\Delta E_{H}$ and $\Delta E_{L}$, we can see that when $\Omega>0$ ($\Omega<0$), as $N$ increases, $\Delta E_{H}$ decreases (increases) and $\Delta E_{L}$ increases (decreases). Here we emphasize that when $N\rightarrow\infty$, $\Delta E_{H}\rightarrow\Delta E_{\frac{N}{2}\rightarrow\frac{N}{2}-1}$ and $\Delta E_{L}\rightarrow\Delta E_{-\frac{N}{2}+1\rightarrow-\frac{N}{2}}$ and both of them are independent of $N$.
On the other hand, when $\Omega>0$ ($\Omega<0$), as $|\Omega|$ increases, $\Delta E_{H}$ decreases (increases) and $\Delta E_{L}$ increases (decreases), and $\Delta E_{L}>\Delta E_{H}$ ($\Delta E_{L}<\Delta E_{H}$) (see Fig. 2).

Defining the ladder operator $L_{M}=|M-1\rangle\langle M|$, the system's dynamical equation Eq. (\ref{EQ}) can be expressed as \cite{Gross1982,Higgins2014}:
\begin{equation}\label{rhos}
\begin{split}
\frac{d\rho_{s}(t)}{dt}=-i[H_{e},\rho_{s}(t)]+&\sum_{M}\Gamma_{M}\biggl(N(\omega_{M})D[L^{\dag}_{M}]\rho_{s}(t)\\
+&\bigl(N(\omega_{M})+1)\bigr)D[L_{M}]\rho_{s}(t)\biggr),
\end{split}
\end{equation}
with $D[A]\rho=A\rho A^{\dag}-\frac{1}{2}\{A^{\dag}A,\rho\}$, $\Gamma_{M}=\frac{4\omega_{M}^{3}}{3}(J-M+1)(J+M)$. Here, we let $|d|^{2}=1$.
If we ignore the interactions between the adjacent atoms ($\Omega=0$), the master equation can be simplified as:
\begin{equation}\label{rh}
\begin{split}
\frac{d\rho_{s}(t)}{dt}=-i[H_{s},\rho_{s}(t)]&+\Gamma_{0}\biggl(N(\omega_{A})D[J_{+}]\rho_{s}(t)\\
&+\bigl(N(\omega_{A})+1)\bigr)D[J_{-}]\rho_{s}(t)\biggr)
\end{split}
\end{equation}
with $\Gamma_{0}=4\omega_{A}^{3}/3$.

It is easy to show that the equilibrium state $\rho_{s}(T)=Z^{-1}\sum_{M}e^{-E_{M}/T}|M\rangle\langle M|$, with $Z=\sum_{M}e^{-E_{M}/T}$, is a fixed point of Eqs. (\ref{rhos}) and (\ref{rh}), i.e., any symmetrical initial state of the probe would arrive at thermal equilibrium with the bath eventually. The problem now goes down to solving Eqs. (\ref{rhos}) and (\ref{rh}). Based on this we can obtain the QFI at any time and further analyze the effects of the structure $\Omega$ and the particle number $N$ of the probe on the thermometry precision. This will be done in the next section, and we would write $\rho$ instead of $\rho_{s}$ for convenience below.

Finally, it should be noted that if we consider a permutation symmetric spin-1/2 chain with the periodic boundary condition, in which the interaction between the spins is intrinsic and is not induced by the environment, as a probe to detect the temperature of a bosonic bath, the energy level structure, the dynamical process (Eqs. (\ref{rhos}) and (\ref{rh})) and the results about the thermometry obtained in the  next section are also valid.

\section{effects of the structure $\Omega$ and the particle number on the thermometry precision}
In this section we apply the quantum estimation theory to estimate the temperature of a bath. An estimation procedure always consists of the following steps: First we send the probe initialized in a quantum state $\rho(0)$ through a sample, which undergoes an evolution depending on some parameter $\theta$; afterwards, we subject the probe to a general quantum measurement, described by a POVM, which outputs measurement results $x$; finally, we should choose an unbiased estimator $\hat{\theta}$ to process the data and infer the value of the unknown parameter $\theta$, and the unbiased estimator $\hat{\theta}$ satisfies $\langle\hat{\theta}\rangle$=$\theta$. This scheme describes not only the quantum estimation tasks but also the classical ones.

The standard deviation of this estimator, i.e., $\Delta\hat{\theta}=\sqrt{Var(\hat{\theta})}$, quantifies the error on estimation of $\theta$. The quantum Cram\'{e}r-Rao bound sets a lower bound on this error as follows:
\begin{equation}\label{crb}
    \Delta\hat{\theta}\geq\frac{1}{\sqrt{\nu Q\bigl(\rho(\theta;t)\bigr)}},
\end{equation}
where $\nu$ is the number of independent experimental repetitions, and $Q(\rho(\theta;t))$ is the quantum Fisher information (QFI) associated with the parameter $\theta$, which is given by:
\begin{equation}\label{qfi}
    Q\bigl(\rho(\theta;t)\bigr)=\mathrm{Tr}\bigl[\rho(\theta;t) L^{2}_{\rho(\theta;t)}\bigr].
\end{equation}
And the symmetric logarithmic derivative $L_{\rho(\theta;t)}$ in the above equation is defined as:
\begin{equation}\label{L}
    \frac{d\rho(\theta;t)}{d\theta}\equiv\frac{1}{2}\bigl(\rho(\theta;t) L_{\rho(\theta;t)}+L_{\rho(\theta;t)}\rho(\theta;t)\bigr).
\end{equation}
Writing $\rho(\theta;t)$ in its spectral decomposition as $\rho(\theta;t)=\sum_{i}p_{i}|\psi_{i}\rangle\langle\psi_{i}|$, one can obtain:
\begin{equation}\label{qqfi}
    Q(\rho(\theta;t))=2\sum_{j,k}\frac{1}{p_{j}+p_{k}}\Big\vert\langle\psi_{j}|\frac{d\rho(\theta;t)}{d\theta}|\psi_{k}\rangle\Big\vert^{2}.
\end{equation}

The computation of the QFI is in general hard since the diagonalization of $\rho(\theta;t)$ is required. However, there exist several upper bounds on the Fisher information \cite{Escher2011,Demkowicz2012}. Nevertheless, for the special initial state we choose, we can easily calculate the
QFI with respect to parameter $T$ at arbitrary time $t$.

In this paper we consider the GHZ-like state $|\psi\rangle=\cos\phi|g,g\cdots g\rangle+\sin\phi|e,e\cdots e\rangle$, ($\phi\in[0,~\pi/2]$), as our initial state of the probe. According to the definition of the Dicke state (Eq. (\ref{Dicke state})), the GHZ-like state can be expressed as $|\psi\rangle=\cos\phi|-J\rangle+\sin\phi|J\rangle$ ($M=-J$ and $J$) and the state at time $t$ can be written as:
\begin{equation}\label{rhoot}
\begin{split}
   \rho(T;t;\phi)&=\rho_{-J,J}(T;t;\phi)|-J\rangle\langle J|+\rho_{J,-J}(T;t;\phi)|J\rangle\langle-J|\\
   &+\sum_{M=-J}^{J}\rho_{M,M}(T;t;\phi)|M\rangle\langle M|,
\end{split}
\end{equation}
where $\rho_{M,M}(T;t;\phi)$ are the diagonal elements of the probe state $\rho$ at time $t$, and $\rho_{-J,J}(T;t;\phi)$ and $\rho_{J,-J}(T;t;\phi)$ are the two nondiagonal elements of $\rho$ at time $t$. This form of the density operator $\rho(T;t;\phi)$ can be diagonalized because it is a $2\times2$ block diagonal.

And then we can calculate the dynamical QFI by putting Eq. (\ref{rhoot}) into Eq. (\ref{qqfi}) associated with the parameter $T$ as follows:
\begin{widetext}
\begin{equation}\label{QFI}
\begin{split}
    &Q_{d}(\rho(T;t;\phi)) \\
    &=\sum_{j,k=+,-}\frac{2}{p_{j}+p_{k}}\Big\vert a^{*}_{j}a_{k}\frac{d\rho_{-J,-J}(T;t;\phi)}{dT}
+a^{*}_{j}b_{k}\frac{d\rho_{-J,J}(T;t;\phi)}{dT}+b^{*}_{j}a_{k}\frac{d\rho_{J,-J}(T;t;\phi)}{dT}+b^{*}_{j}b_{k}\frac{d\rho_{J,J}(T;t;\phi)}{dT}\Big\vert^{2} \\
             &+\sum_{M=-J+1}^{J-1}\frac{1}{\rho_{M,M}(T;t;\phi)}\Big\vert\frac{d\rho_{M,M}(T;t;\phi)}{dT}\Big\vert^{2},
\end{split}
\end{equation}
\end{widetext}
where $a_{-}=-\frac{\rho_{-J,J}(T;t;\phi)}{\chi_{-}}$, $a_{+}=\frac{\rho_{J,J}(T;t;\phi)-p_{+}}{\chi_{+}}$,
$b_{-}=\frac{\rho_{-J,-J}(T;t;\phi)-p_{-}}{\chi_{-}}$,
$b_{+}=-\frac{\rho_{J,-J}(T;t;\phi)}{\chi_{+}}$,
with $p_{\pm}=\frac{1}{2}((\rho_{-J,-J}(T;t;\phi)+\rho_{J,J}(T;t;\phi))\pm\eta)$, $\eta=\sqrt{4|\rho_{-J,J}(T;t;\phi)|^{2}+(\rho_{-J,-J}(T;t;\phi)-\rho_{J,J}(T;t;\phi))^{2}}$ and
$\chi_{\pm}=\sqrt{|\rho_{-J,J}(T;t;\phi)|^{2}+(\rho_{\pm J,\pm J}(T;t;\phi)-p_{\pm})^{2}}$.
And from Eq. (\ref{QFI}) we can see that the nondiagonal elements $\rho_{J,-J}(T;t;\phi)$ and $\rho_{-J,J}(T;t;\phi)$, associated with quantum coherence, affect the dynamical QFI.

On the other hand, for temperature estimation on a thermal equilibrium state $\rho(T)$, the QFI is analytically given by \cite{Zanardi2008,Haupt2014}:
\begin{equation}\label{qqqfi}
    Q_{e}\bigl(\rho(T)\bigr)=\frac{\Delta H_{e}^{2}}{T^{4}},
\end{equation}
where
\begin{equation}\label{DATAH}
\begin{split}
\Delta H_{e}^{2}\equiv& \mathrm{Tr}\bigl(H_{e}^{2}\rho(T)\bigr)-\biggl(\mathrm{Tr}\bigl(H_{e}\rho(T)\bigr)\biggr)^{2}\\
               =&\frac{1}{Z}\sum_{M}E^{2}_{M}e^{\frac{-E_{M}}{T}}-\frac{1}{Z^{2}}\biggl(\sum_{M}E_{M}e^{\frac{-E_{M}}{T}}\biggr)^{2},
\end{split}
\end{equation}
with $\rho(T)=Z^{-1}\sum_{M}e^{-E_{M}/T}|M\rangle\langle M|$ and $Z=\sum_{M}e^{-E_{M}/T}$ mentioned below Eq. (\ref{rh}). In light of Eq. (\ref{qqqfi}), we can see that the maximization of the QFI at a given $T$ is equivalent to the maximization of the energy variance at thermal equilibrium, i.e., more levels are populated.
And from Eq. (\ref{QFI}) it can be verified that:
\begin{equation}\label{de}
   \lim_{t\rightarrow\infty}Q_{d}\bigl(\rho(T;t;\phi))=Q_{e}(\rho(T)\bigr).
\end{equation}
Here it should be noted that $t\rightarrow\infty$ in Eq. (\ref{de}) just means that finally the probe should be in equilibrium with the bath, and from our numerical calculation we find that in most cases the probe can be approximately in equilibrium with the bath in a finite time.

In what follows, we will analyze the effects of the structure $\Omega$ and the particle number $N$ of the probe on the thermometry precision in two complementary scenarios, the thermal equilibrium thermometry and the dynamical thermometry.

\subsection{Thermal equilibrium thermometry}
First, we consider the thermal equilibrium thermometry, where the measurement is taken when the probe reaches thermal equilibrium with the bath, so it is irrespective of what the symmetrical initial state of the probe we choose. According to Eqs. (\ref{qqqfi}) and (\ref{DATAH}) and through our calculations, we find that for the thermal equilibrium thermometry, the ferromagnetic structure ($\Omega<0$) has an advantage over the non-structure case ($\Omega=0$) in the thermometry precision while the antiferromagnetic structure ($\Omega>0$) does not. Besides, we find that at very low temperature, the behavior of the equilibrated QFI is almost independent of the particle number $N$. While as the temperature increases, the particle number $N$ becomes somewhat related to the value of the equilibrated QFI.
\begin{center}
\includegraphics[width=8cm]{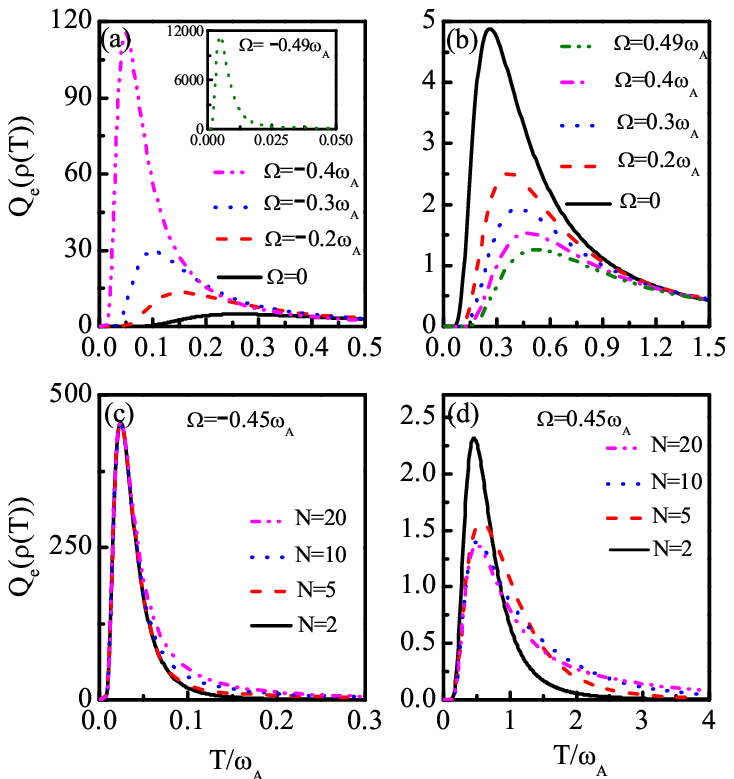}
\parbox{8cm}{\small{FIG. 3.} (Color online) (a) and (b) quantum Fisher information, $Q_{e}(\rho(T))$, as a function of the bath temperature $T$ for different coupling strength with $N=20$ and for the ferromagnetic structure and the antiferromagnetic structure, respectively, and the inset of Fig. 3.(a) is for $\Omega=-0.49\omega_{A}$ with $N=20$; (c) and (d) quantum Fisher information, $Q_{e}(\rho(T))$, as a function of the bath temperature $T$ for different particle numbers $N=2,~5,~10,~20$ and for the ferromagnetic structure $\Omega=-0.45\omega_{A}$ and the antiferromagnetic structure $\Omega=0.45\omega_{A}$, respectively. $k_{B}=\hbar=1$.}
\end{center}

As an example, in Figs. 3(a) and (b), we plot the equilibrated QFI (Eq. (\ref{qqqfi})), as a function of the bath temperature $T$ for different coupling strength with $N=20$ in the case of the ferromagnetic structure and the antiferromagnetic structure, respectively. We can see from Figs. 3(a) and (b) that as the coupling strength $\Omega$ varies from $-0.49\omega_{A}\rightarrow 0.49\omega_{A}$, the optimal temperature $T_{opt}$ becomes higher and the value of its corresponding equilibrated QFI becomes smaller gradually. It can be clearly seen that for the ferromagnetic structure ($\Omega<0$), the thermometry precision is always higher than that of the non-structure case ($\Omega=0$) (see Fig. 3(a)), for example, when the coupling strength $\Omega=-0.49\omega_{A}$, the value of the equilibrated QFI is very large and $T_{opt}$ is very low. While for the antiferromagnetic structure ($\Omega>0$), the thermometry precision is lower than that of the non-structure case ($\Omega=0$) (see Fig. 3(b)). It can be concluded that for the thermal equilibrium thermometry, the ferromagnetic structure has an advantage over the non-structure case ($\Omega=0$) in the attainable thermometry precision, i.e., it can measure a lower temperature of the bath with an improved precision than the non-structure probe.

This can be understood as follows. For the thermal equilibrium thermometry and at low temperature, the atoms are mainly distributed in the lower energy levels, so the distribution of the lower energy levels plays a leading role in the thermometry precision. Besides, we know that the smaller the energy difference, the more sensitive the probe to the thermal fluctuation, because even at a very low temperature, almost all the lower energy levels can be populated, that is, the energy variance $\Delta H_{e}^{2}$ is large, and hence the resulting equilibrated QFI is large, and vice versa. As a result, when $\Omega<0$, the lower energy difference is smaller than that of the non-structure case ($\Omega=0$) (see Fig. 2), so the value of the equilibrated QFI is larger than that of the non-structure case. Moreover, for the ferromagnetic structure ($\Omega<0$), the larger the absolute value of the coupling strength $\Omega$, the smaller the lower energy difference $\Delta E_{L}$, which has been analyzed in Sec. \uppercase\expandafter{\romannumeral2}, so the lower the $T_{opt}$, and the larger the value of the equilibrated QFI at its corresponding $T_{opt}$. Note that at $T_{opt}$, the transition frequencies of the probe, corresponding to the lower energy differences $\Delta E_{L}$ are close to resonance with the characteristic frequency of the thermal fluctuation of the bath \cite{Correa2015}. In contrast, when $\Omega>0$, the lower energy difference is larger than that of the non-structure case ($\Omega=0$) (see Fig. 2), so the value of the equilibrated QFI is smaller than that of the non-structure case. Moreover, for the antiferromagnetic structure ($\Omega>0$), the larger the coupling strength $\Omega$, the larger the lower energy difference $\Delta E_{L}$, so the higher the $T_{opt}$, and the smaller the value of the equilibrated QFI at its corresponding $T_{opt}$.

Next, in Figs. 3(c) and (d), we plot the equilibrated QFI (Eq. (\ref{qqqfi})), as a function of the bath temperature $T$ for different particle number $N$ in the case of the ferromagnetic structure $\Omega=-0.45\omega_{A}$ and the antiferromagnetic structure $\Omega=0.45\omega_{A}$, respectively. We can see from Fig. 3(c) that at very low temperature ($T\in(0,~0.05)$), the behaviors of the equilibrated QFI for the ferromagnetic structure are almost the same for different particle number $N$, while as the temperature increases, slight difference appears among them. Specifically, the larger the particle number $N$, the larger the value of the equilibrated QFI. On the contrary, for the antiferromagnetic structure, from Fig. 3(d) we can see that at relatively low temperature ($T\in(0,~0.7)$), the larger the particle number $N$, the smaller the value of the equilibrated QFI; while at relatively high temperature ($T>2.5$), the larger the particle number $N$, the larger the value of the equilibrated QFI.

This can be illustrated from the point of the energy level structure (refer to Fig. 2 and the analysis in Sec. \uppercase\expandafter{\romannumeral2}). Specifically, when $\Omega<0$, the lower energy difference $\Delta E_{L}$ is smaller than the higher energy difference $\Delta E_{H}$ and the populations of the lower energy levels are greater than that of the higher energy levels. So the lower energy difference plays a leading role in the thermal equilibrium thermometry. As the particle number $N$ increases, the lower energy difference $\Delta E_{L}$ becomes smaller, so the value of the equilibrated QFI increases. Note that at a very low temperature, the atoms are almost distributed in the ground level and the first excited level and due to the fact that the energy difference between the ground level and the first excited level is independent of the particle number $N$, so the equilibrated QFI is almost independent of the particle number $N$, which can be seen in Fig. 3(c). On the contrary, when $\Omega>0$, the lower energy difference $\Delta E_{L}$ is larger than the higher energy difference $\Delta E_{H}$ but the populations of the lower energy levels are greater than that of the higher energy levels. In this case things become complicated and there might be a trade-off between the population and the energy difference. So in this regime ($\Omega>0$), the equilibrated QFI is not monotonous with respect to $N$ in the entire temperature range. Specifically, for relatively low temperature, the higher energy levels get almost unpopulated, so the lower energy difference $\Delta E_{L}$ plays a leading role in the thermometry precision. As a result, the larger the particle number $N$, the larger the $\Delta E_{L}$, and thus the smaller the value of the equilibrated QFI; while for relatively high temperature, the higher energy levels start to get populated, and the higher energy difference $\Delta E_{H}$ is smaller than the lower energy difference $\Delta E_{L}$, so the higher energy difference $\Delta E_{H}$ may play an indispensable role in the thermometry precision. As a result, the larger the particle number $N$, the smaller the $\Delta E_{H}$, and thus the larger the value of the equilibrated QFI, which can be seen in Fig. 3(d). Here we emphasize that for each coupling strength $\Omega\in(-0.5\omega_{A},~0.5\omega_{A}$), when $N \rightarrow\infty$, i.e., in the thermodynamic limit, the equilibrated QFI is independent of $N$, because when $N \rightarrow\infty$, both $\Delta E_{L}$ and $\Delta E_{H}$ are independent of $N$ (see Sec. \uppercase\expandafter{\romannumeral2}).

Besides, we also investigate the effects of the coupling strength $\Omega$ and the particle number $N$ on the time $t_{e}$ that the probe needed to arrive at equilibrium with the bath. In this paper we numerically calculate $t_{e}$ as following. For each bath temperature $T$, we first give the analytical expression of the thermal equilibrated QFI, $Q_{e}(\rho(T))$, and then we calculate the dynamical QFI, $Q_{d}(\rho(T;t;\phi))$, of the probe state at time $t$, and in our numerical calculations we define $t_{e}$ as the shortest time which satisfies $|Q_{d}(\rho(T;t_{e};\phi))- Q_{e}(\rho(T))|<10^{-12}$, and in this case we suppose that the probe is approximately in equilibrium with the bath at temperature $T$. And
through our numerical calculations, we find that the closer the coupling strength $\Omega$ is to the energy level crossing ($\Omega=\pm0.5\omega_{A}$), the slower the probe evolves to the thermal equilibrium state no matter what the symmetrical initial state of the probe is. Fortunately, we can reduce the time $t_{e}$ by increasing the particle number $N$ of the probe, that is, as the particle number $N$ increases, $t_{e}$ becomes shorter. In Fig. 4, we plot $t_{e}$, at the optimal temperature $T_{opt}$, as a function of the particle number $N$ for (a) $\Omega=-0.3\omega_{A}$, (b) $\Omega=-0.4\omega_{A}$ and (c) $\Omega=-0.45\omega_{A}$ and we choose the ground state ($\phi=0$) as the probe initial state for simplicity. Here we emphasize that some other forms of the symmetrical initial state are also available for calculating $t_{e}$, however, the ground state is relatively quicker to arrive at thermal equilibrium with the bath than others. We can see from Fig. 4 that for fixed particle number $N$, the larger the absolute value of the coupling strength $\Omega$, the longer the time $t_{e}$. However, for fixed coupling strength $\Omega$, the larger the particle number $N$, the shorter the time $t_{e}$. In particular, we can see from Fig. 4 that the time $t_{e}$ is approximately scaled by $1/N$ for any coupling strength $\Omega$. This is because that the decay rates appearing in Eq. (\ref{rhos}) are $\Gamma_{M}=\frac{4\omega_{M}^{3}}{3}(J-M+1)(J+M)$, where the minimal value of $(J-M+1)(J+M)$ $\sim N$ (when $J=N/2,~M=J$ or $-J$).
\begin{widetext}
\begin{center}
\includegraphics[width=15cm]{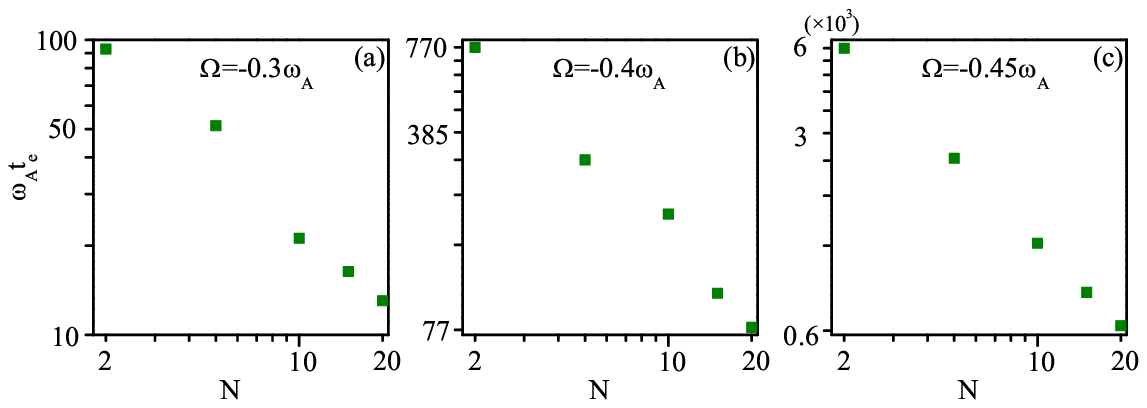}
\parbox{15cm}{\small{FIG. 4.} (Color online) The time that the probe needed to reach thermal equilibrium with the bath, $t_{e}$, at the optimal temperature $T_{opt}$, as a function of the particle number $N$ for (a) $\Omega=-0.3\omega_{A}$, (b) $\Omega=-0.4\omega_{A}$ and (c) $\Omega=-0.45\omega_{A}$. $\phi=0$.}
\end{center}
\end{widetext}
\begin{widetext}
\begin{center}
\includegraphics[width=15cm]{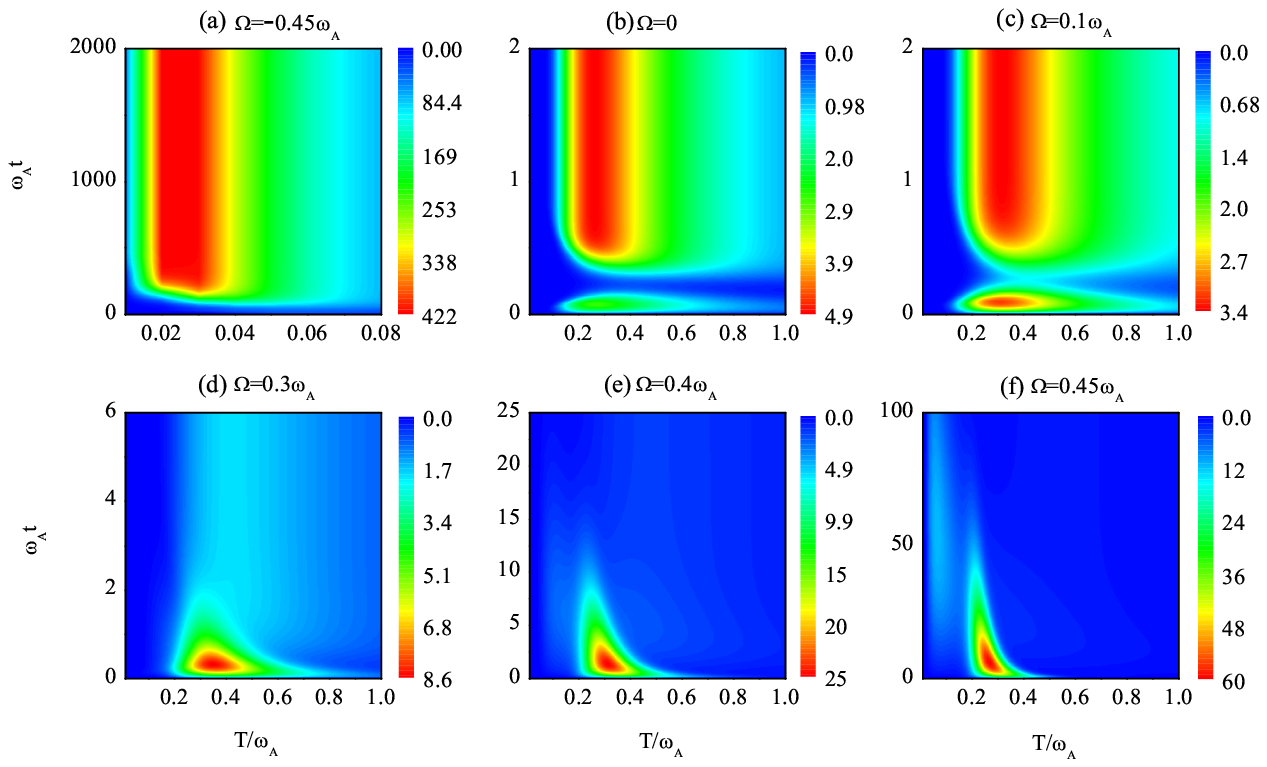}
\parbox{15cm}{\small{FIG. 5.} (Color online) Quantum Fisher information, $Q_{d}(\rho(T;t))$, as functions of the bath temperature $T$ and the measurement time $t$ for different coupling strengths $\Omega=-0.45\omega_{A},~0,~0.1\omega_{A},~0.3\omega_{A},~0.4\omega_{A},~0.45\omega_{A}$ with $N=20$. $\phi=\pi/4$, $k_{B}=\hbar=1$.}
\end{center}
\end{widetext}

\subsection{Dynamical thermometry}
All the previous analyses are focused on the thermal equilibrium thermometry. In practice, however, one may have to read out the temperature before achieving full thermalization due to some constraint, for example, the probe is very hard to reach thermal equilibrium with the bath, i.e., $t_{e}$ is very long. Now, let us analyze the effects of the structure $\Omega$ and the particle number $N$ of the probe on the dynamical thermometry (see Eqs. (\ref{rhos}) and (\ref{rh})) where the measurement is taken before the thermal equilibrium is reached.
In order to maximize the dynamical QFI, $Q_{d}(\rho(T;t;\phi))$ (Eq. (\ref{QFI})), we numerically optimize the GHZ-like initial state over $\phi$ and find that the optimal initial state is the standard GHZ state, i.e., $\phi=\pi/4$. So we will use the standard GHZ state as our initial state of the probe for the dynamical thermometry in the following. And it is noted that for simplicity we will omit $\phi$ in the expression of $Q_{d}(\rho(T;t;\phi))$, i.e., we will use $Q_{d}(\rho(T;t))$ to represent the QFI for the standard GHZ state in the following. Based on Eq. (\ref{QFI}) and through our calculations, we find that for the dynamical thermometry, the antiferromagnetic structure ($\Omega>0$) plays a distinctive role in the thermometry precision, that is, it can make the dynamical QFI larger than the equilibrated QFI, and on this condition, increasing $N$ in a limited range can increase the value of the dynamical QFI. Moreover, the best precision achieved in the case of the antiferromagnetic structure can be higher than that in the non-structure case.

While for the ferromagnetic structure and the non-structure, through our numerical calculations, we find that they can not make the dynamical QFI larger than the equilibrated QFI. As an example, we plot Figs. 5(a) and (b) to show the behaviors of the QFI as functions of the bath temperature $T$ and the measurement time $t$ with $N=20$ in the ferromagnetic and the non-structure cases, respectively. We can see from Figs. 5(a) and (b) that for both the ferromagnetic structure and the non-structure, the QFI would arrive at its largest value when the probe is equilibrated with the bath.

In contrast, in Figs. 5(c-f) we plot the QFI as functions of the bath temperature $T$ and the measurement time $t$ for different coupling strengths $\Omega=0.1\omega_{A},~0.3\omega_{A},~0.4\omega_{A},~0.45\omega_{A}$ in the case of the antiferromagnetic structure with $N=20$. We can see that for small coupling strength, for example, $\Omega=0.1\omega_{A}$, the QFI arrives at its largest value when the probe is equilibrated with the bath. But for relatively larger coupling strengths, i.e., $\Omega=0.3\omega_{A},~0.4\omega_{A},~0.45\omega_{A}$, the largest value of the QFI appears in the dynamical process, rather than in the equilibrium state. Besides, we can see from Figs. 5(d-f) that as the coupling strength increases, the optimal QFI, $Q_{d}(\rho(T_{opt},t_{opt}))$, which has the largest value in the entire $T-t$ parameter space (the reddest point in Fig. 5), becomes increasingly larger.

Furthermore, in Fig. 6(a), we plot the time optimized QFI, i.e., $Q_{M}(\rho(T))\equiv\underset{t}{max}Q_{d}(\rho(T;t))$, as a function of the bath temperature $T$ for different coupling strengths $\Omega=0,~0.3\omega_{A},~0.4\omega_{A},~0.45\omega_{A},~0.49\omega_{A}$ with $N=20$. The maximization above is carried out over all the measurement time $t$ during the evolution of the probe, at which the dynamical QFI $Q_{d}(\rho(T;t))$ achieves its largest value. The inset of Fig. 6(a) is for $\Omega=0.49\omega_{A}$. We can see that in the case of the antiferromagnetic structure ($\Omega>0$), as the coupling strength $\Omega$ increases, the optimal QFI is monotonically increasing and $T_{opt}$ is generally shifting towards the low temperature region slightly. That is, the larger the coupling strength, the lower the $T_{opt}$, and the larger the optimal QFI. And we can also see that the optimal QFI obtained in the case of the antiferromagnetic structure can be much larger than that in the non-structure case.

The above phenomena can be interpreted as follows. Different from the thermal equilibrium thermometry, for the dynamical thermometry, all the energy levels would contribute to the thermometry precision, not just the lower energy levels. This is because that the initial state, GHZ state, is equally distributed in the highest excited level and the ground level, so the transitions between the adjacent higher energy levels are bound to happen during the evolution. For $\Omega=0$, i.e., the non-structure case, all the energy levels are equally spaced, so the dynamical QFI can not be larger than the equilibrated QFI. For $\Omega<0$, the lower energy difference $\Delta E_{L}$ is smaller than the higher energy difference $\Delta E_{H}$, and the populations of the lower energy levels are greater than that of the higher energy levels during the evolution, so the contributions of the higher energy levels to the precision are less than that of the lower energy levels. As a result, the dynamical QFI also can not be larger than the equilibrated QFI. While for $\Omega>0$, although the populations of the higher energy levels are still smaller than that of the lower energy levels during the time evolution, the higher energy difference $\Delta E_{H}$ is smaller than the lower energy difference $\Delta E_{L}$. So there exists a competition in the contribution to the thermometry precision between the population and the energy difference, and only when the higher energy difference $\Delta E_{H}$ is far smaller than the lower energy difference $\Delta E_{L}$, the dynamical QFI can be larger than the equilibrated QFI. And on this condition, the larger the coupling strength, the smaller the higher energy difference $\Delta E_{H}$, so the lower the $T_{opt}$, and the larger the value of the optimal QFI. Note that at $T_{opt}$, the transition frequencies of the probe, corresponding to the higher energy differences $\Delta E_{H}$ are close to resonance with the characteristic frequency of the thermal fluctuation of the bath. In addition, the optimal QFI obtained in the case of the antiferromagnetic structure can be much larger than that in the non-structure case because when the coupling strength $\Omega$ increases to a certain value, the higher energy difference can be much smaller than the equally spaced energy difference of the non-structure case.
\begin{center}
\includegraphics[width=8cm]{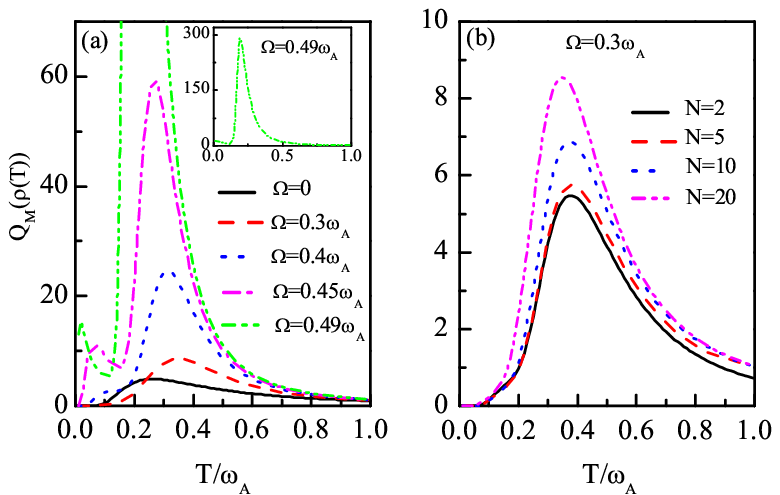}
\parbox{8cm}{\small{FIG. 6.} (Color online) Time optimized QFI, $Q_{M}(\rho(T))$, as  a function of the bath temperature $T$ for (a) different coupling strengths $\Omega=0,~0.3\omega_{A},~0.4\omega_{A},~0.45\omega_{A},~0.49\omega_{A}$ with $N=20$, and the inset gives the complete behavior of $Q_{M}(\rho(T))$ for $\Omega=0.49\omega_{A}$; (b) different particle numbers $N=2,~5,~10,~20$ with $\Omega=0.3\omega_{A}$. $\phi=\pi/4$, $k_{B}=\hbar=1$. }
\end{center}

Next, in Fig. 6(b) we plot the time optimized QFI, $Q_{M}(\rho(T))$, as a function of the bath temperature $T$ for different particle numbers $N = 2,~ 5,~ 10,~ 20$ with $\Omega=0.3\omega_{A}$. We can see from Fig. 6(b) that as the particle number $N$ increases, the value of the optimal QFI becomes larger. This is because that as the particle number $N$ increases, the higher energy difference $\Delta E_{H}$, which plays a leading role in this case, becomes gradually smaller, so the optimal QFI becomes increasingly larger. And through our numerical calculations we find that when $N\rightarrow\infty$, i.e., in the thermodynamic limit, the behavior of the dynamical QFI is independent of $N$.
\begin{widetext}
\begin{center}
\includegraphics[width=15cm]{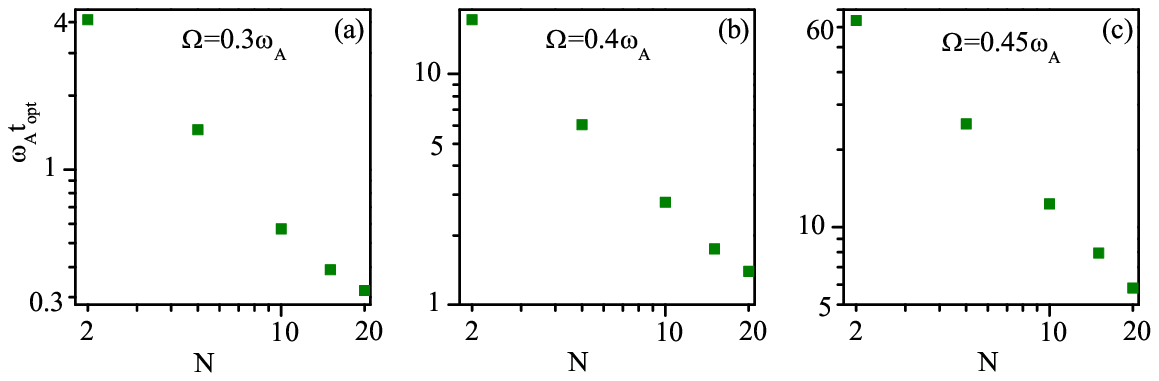}
\parbox{15cm}{\small{FIG. 7.} (Color online) Optimal measurement time $t_{opt}$, corresponding to $Q_{d}(\rho(T_{opt},t_{opt}))$, as a function of the particle number $N$ for (a) $\Omega=0.3\omega_{A}$, (b) $\Omega=0.4\omega_{A}$ and (c) $\Omega=0.45\omega_{A}$. $\phi=\pi/4$.}
\end{center}
\end{widetext}

Similar to the thermal equilibrium thermometry, in the dynamical thermometry, the closer the coupling strength $\Omega$ is to the energy level crossing ($\Omega=0.5\omega_{A}$), the slower the probe evolves and the longer the optimal measurement time $t_{opt}$ is, corresponding to the optimal QFI, $Q_{d}(\rho(T_{opt},t_{opt}))$. Fortunately, we can also reduce $t_{opt}$ by increasing the particle number $N$ of the probe. In Fig. 7 we plot the optimal measurement time $t_{opt}$ as a function of the particle number $N$ for (a) $\Omega=0.3\omega_{A}$, (b) $\Omega=0.4\omega_{A}$ and (c) $\Omega=0.45\omega_{A}$. We can see that for fixed particle number $N$, the larger the coupling strength $\Omega$, the longer the optimal measurement time $t_{opt}$. However, for fixed coupling strength $\Omega$, the larger the particle number $N$, the shorter the optimal measurement time $t_{opt}$. And we can see from Fig. 7 that $t_{opt}$ is also approximately scaled by $1/N$.

\section{Conclusions}
In summary, we have investigated the effects of the structure $\Omega$ (strength of the dipole-dipole interaction between adjacent atoms) and the particle number $N$ of a ring-structure probe on the thermometry precision of an electromagnetic field (bath) in two complementary scenarios, i.e., the thermal equilibrium thermometry and the dynamical thermometry. We have calculated the quantum Fisher information (QFI) of the probe for our model at any time and then analyzed what roles the structure and the particle number of the probe would play in the temperature estimation. We have found that for the thermal equilibrium thermometry, the ferromagnetic structure ($\Omega<0$) can measure a lower temperature of the bath with a higher precision compared with the non-structure probe ($\Omega=0$). More accurately, as the absolute value of the coupling strength $\Omega$ increases, the optimal temperature becomes lower and the value of the corresponding equilibrated QFI of the probe becomes larger. However, the probe would take a longer time to be equilibrated with the bath. Fortunately, we can reduce it by increasing the particle number $N$ of the probe. Moreover, for the ferromagnetic structure, increasing $N$ in a limited range can also improve the thermometry precision more or less especially for a relatively higher temperature. In contrast, for the dynamical thermometry, the antiferromagnetic structure ($\Omega>0$) would play an important role. Specifically, when the coupling strength $\Omega$ increases to a certain value, the QFI of the probe in the dynamical process can be much larger than that in equilibrium with the bath, which is somewhat counterintuitive. Moreover, the best precision achieved in the case of the antiferromagnetic structure can be much higher than that in the non-structure case. Specifically, the larger the coupling strength, the lower the optimal temperature and the larger the optimal QFI. But the optimal measurement time becomes longer. Similarly, we can reduce it by increasing the particle number $N$ of the probe. Additionally, increasing $N$ in a limited range can also increase the value of the dynamical QFI.

While in this paper we have not discussed the effect of the quantum correlation on the thermometry, and actually we have tried various initial states during our research besides the GHZ-like state to study their dynamical thermometry, including the ground state and the excited state which have neither the classical correlation nor the quantum correlation, the maximally mixed state which has classical correlation, the superposition state composed of the highest excited state $|N/2\rangle$ and the second highest excited state $|N/2-1\rangle$ and some other forms of superposition state which have both classical and quantum correlations. And we have found that among all the initial states we considered, the standard GHZ state performs the best for the dynamical thermometry. So we have finally chosen the standard GHZ state as our initial state of the probe for the dynamical thermometry. But the exact mechanism of how the coherence can promote the dynamical thermometry and what is the potential role played by quantumness in thermometry are complicated and still unknown for us, which deserves a deep investigation in our further work.
On the other hand, we have just investigated a relatively simple energy level structure of our model which does not involve the energy level crossing, but when the parameter $\Omega$ extends the regime we considered in this paper, there would be some energy level crossings and things become complicated. We have found that at these points (energy level crossing) the evolution speed of the probe would be reduced significantly, which is associated with the quantum phase transition. And we are very curious about what would happen for the thermometry around these phase transition points and what is the relationship between quantum phase transition and thermometry, which would be investigated in our future work.

\begin{acknowledgments}
This work was supported by the National Natural Science
Foundation of China (Grants No. 11274043, 11375025).
\end{acknowledgments}

\appendix

\section{The derivation of the master equation}
For the model we consider in this paper, i.e., Eqs. (\ref{HS}-\ref{HI}) in the main text, after performing the standard Born-Markov approximation and taking the trace over the environment, the starting point for our derivation is \cite{Gross1982,Higgins2014,Breuer2002}
\begin{equation}\label{}
\begin{split}
  &\frac{d\rho_{s}(t)}{dt}=-i[H_{s},\rho_{s}(t)] \\
  &+\sum_{\omega}\sum_{m,n}
  \bigl[\Gamma_{mn}(\omega)\bigl(\sigma^{n}_{-}\rho_{s}(t)\sigma^{m}_{+}-\sigma^{m}_{+}\sigma^{n}_{-}\rho_{s}(t)\bigr)+h.c.\bigr],
\end{split}
\end{equation}
where h.c. denotes the Hermitian conjugate. $\Gamma_{mn}(\omega)=\int_{0}^{\infty}dse^{i\omega s}\mathrm{Tr}_{B}[\textbf{d}^{\ast}\cdot\hat{\textbf{E}}(\textbf{r}_{m},s)\textbf{d}\cdot\hat{\textbf{E}}(\textbf{r}_{n},0)\rho_{B}(T)]$ is the spectral correlation tensor, $\rho_{B}(T)=\frac{1}{Z}\exp[-H_{B}/T]$ is the thermal state of the environment with the partition function $Z=\mathrm{Tr}[\exp[{-H_{B}/T}]]$ and $T$ being the temperature (here we let the Boltzmann constant $k_{B}=1$). In this case, the spectral correlation tensor $\Gamma_{mn}(\omega)$ can be expressed as \cite{Gross1982,Higgins2014}
\begin{widetext}
\begin{equation}\label{}
\Gamma_{mn}(\omega)=\frac{|d|^{2}}{4\pi}\int_{0}^{\infty}d\omega_{k} \kappa(\omega_{k})\omega_{k}^{3}F(\omega_{k} \textbf{r}_{mn})\biggl(\bigl(1+N(\omega_{k})\bigr)e^{i\textbf{k}\cdot\textbf{r}_{mn}}\int_{0}^{\infty}ds e^{-i(\omega_{k}-\omega)s}+N(\omega_{k})e^{-i\textbf{k}\cdot\textbf{r}_{mn}}\int_{0}^{\infty}ds e^{i(\omega_{k}+\omega)s} \biggr),
\end{equation}
\end{widetext}
where $\kappa(\omega)=\sum_{k}|g_{k}|^{2}\delta(\omega-\omega_{k})$ is the spectral density, $F(\omega_{k} \textbf{r}_{mn})$ is a diffraction-type function with the vector $\textbf{r}_{mn}=\textbf{r}_{n}-\textbf{r}_{m}$. Due to $\int_{0}^{\infty}dse^{\pm i\epsilon s}=\pi\delta(\epsilon)\pm iP\frac{1}{\epsilon}$ (where $P$ denotes the Cauchy principal value), the spectral correlation tensor $\Gamma_{mn}(\omega)$ can be divided into two parts $\Gamma_{mn}(\omega)=\gamma_{mn}(\omega)+iS(\omega)$.

The real part $\gamma_{mn}$ is derived from the $\delta$-functions and gives rise to the dissipative dynamics. In this paper we assume that all atomic dipoles are parallel, and perpendicular to the plane defined by the ring. And we are working in the small atomic system limit, where the wavelength of the electromagnetic field is far longer than the size of our probe, i.e., $\omega r_{mn}\approx0$. In this case, $F(\omega_{k} \textbf{r}_{mn})\approx8\pi/3$.
For a flat spectral density, the dissipative rate can be expressed as \cite{Gross1982,Higgins2014}
\begin{equation}
  \gamma_{mn}(\omega)\approx\gamma(\omega)=\frac{4\omega^{3}|d|^{2}}{3}\bigl(1+N(\omega)\bigr),
\end{equation}
where $N(\omega)=(\exp[\omega/T]-1)^{-1}$ is the Planck distribution with the property $N(-\omega)=-(1+N(\omega))$. The dissipative dynamics corresponding to the real part is
\begin{widetext}
\begin{equation}\label{EQ1}
\begin{split}
\bigl(\frac{d\rho_{s}(t)}{dt}\bigr)_{real}=\sum_{\omega>0}\sum^{N}_{m,n}\frac{4\omega^{3}|d|^2}{3}
\biggl(\bigl(1+N(\omega)\bigr)\bigl(\sigma^{n}_{-}\rho_{s}(t)\sigma^{m}_{+}-\frac{1}{2}\big\{\sigma^{m}_{+}\sigma^{n}_{-},\rho_{s}(t)\big\}\bigr)
+N(\omega)\bigl(\sigma^{n}_{+}\rho_{s}(t)\sigma^{m}_{-}-\frac{1}{2}\big\{\sigma^{m}_{-}\sigma^{n}_{+},\rho_{s}(t)\big\}\bigr)\biggr),
\end{split}
\end{equation}
\end{widetext}
where $\{\cdot,\cdot\}$ represents the anticommutator.

We now turn to the imaginary part $S(\omega)$ of the spectral correlation tensor. The $m = n$ terms, for which $F(0) = 8\pi/3$, correspond to the ordinary Lamb shift of individual atom transitions; these can be accounted for by a renormalization of the bare atomic frequency $\omega_{A}$. By contrast, the $m\neq n$ terms correspond to the Van der Waals dipole-dipole interaction induced by the electromagnetic field. For a small atomic system $\omega r_{mn}\ll1$, the Van der Waals dipole-dipole interaction can be described as (here, we neglect the Stark shifts):
\begin{equation}\label{Hdd}
  H_{d}=\sum^{N}_{m>n}\Omega_{mn}(\sigma^{m}_{+}\sigma^{n}_{-}+\sigma^{n}_{-}\sigma^{m}_{+}),
\end{equation}
with the interaction strength $\Omega_{mn}$ given by \cite{Gross1982,Higgins2014}
\begin{equation}
  \Omega_{mn}=\frac{d^{2}}{4\pi r_{mn}^{3}}\biggl[1-\frac{3(\hat{\bm{\epsilon}}_{a}\cdot\textbf{r}_{mn})^{2}}{r_{mn}^{2}}\biggr]\approx\frac{d^{2}}{4\pi r_{mn}^{3}},
\end{equation}
where $\hat{\bm{\epsilon}}_{a}$ is a unit vector parallel to the direction of the dipoles.
Due to the $1/r_{mn}^{3}$ decreasing of the dipole-dipole interaction, the interactions between the adjacent atoms play a dominant role, additionally, for a symmetric geometries, i.e., the ring structure considered in this paper, the interaction strength $\Omega_{mn}:=\Omega$ is a constant, such that the dipole-dipole interaction (Eq. (\ref{Hdd})) reduces to:
\begin{equation}\label{}
  H_{d}=\Omega\sum_{n}(\sigma^{n}_{+}\sigma^{n+1}_{-}+\sigma^{n}_{-}\sigma^{n+1}_{+}),
\end{equation}
where $\Omega=\frac{d^{2}}{4\pi r^{3}}$, $r$ represents the distance between the neighbor atoms, and the periodic boundary condition $\sigma^{N+1}_{\pm}=\sigma^{1}_{\pm}$ is considered. The dynamics corresponding to the imaginary part can be described as \cite{Gross1982,Higgins2014}
\begin{equation}\label{EQ2}
\bigl(\frac{d\rho_{s}(t)}{dt}\bigr)_{imag}=-i[H_{s}+H_{d},\rho_{s}(t)].
\end{equation}
Combining Eqs. (\ref{EQ1}) and (\ref{EQ2}), the dynamics of the probe system can be expressed as
\begin{widetext}
\begin{equation}\label{}
\begin{split}
\frac{d\rho_{s}(t)}{dt}=&-i[H_{s}+H_{d},\rho_{s}(t)]\\
&+\sum_{\omega>0}\sum^{N}_{m,n}\frac{4\omega^{3}|d|^2}{3}
\biggl(\bigl(1+N(\omega)\bigr)\bigl(\sigma^{n}_{-}\rho_{s}(t)\sigma^{m}_{+}-\frac{1}{2}\big\{\sigma^{m}_{+}\sigma^{n}_{-},\rho_{s}(t)\big\}\bigr)
+N(\omega)\bigl(\sigma^{n}_{+}\rho_{s}(t)\sigma^{m}_{-}-\frac{1}{2}\big\{\sigma^{m}_{-}\sigma^{n}_{+},\rho_{s}(t)\big\}\bigr)\biggr).
\end{split}
\end{equation}
\end{widetext}


\begin{thebibliography}{}
\bibitem{Caves1981} C. M. Caves, {Phys. Rev. D} \textbf{23}, 1693 (1981).
\bibitem{McKenzie2002} K. McKenzie, D. A. Shaddock, D. E. McClelland, B. C. Buchler, and P. K. Lam, {Phys. Rev. Lett.} \textbf{88}, 231102 (2002).
\bibitem{Wineland1992} D. J. Wineland, J. J. Bollinger, W. M. Itano, F. L. Moore, and D. J. Heinzen, {Phys. Rev. A} \textbf{46}, R6797 (1992).
\bibitem{Bollinger1996} J. J. Bollinger, W. M. Itano, D. J. Wineland, and D. J. Heinzen, {Phys. Rev. A} \textbf{54}, R4649 (1996).
\bibitem{Holland1993} M. J. Holland and K. Burnett, {Phys. Rev. Lett.} \textbf{71}, 1355 (1993).
\bibitem{Lee2002} H. Lee, P. Kok and J. P. Dowling, {J. Mod. Opt.} \textbf{49}, 2325 (2002).
\bibitem{Valencia2004} A. Valencia, G. Scarcelli, and Y. Shih, {Appl. Phys. Lett.} \textbf{85}, 2655 (2004).
\bibitem{de Burgh2005} M. de Burgh and S. D. Bartlett, {Phys. Rev. A} \textbf{72}, 042301 (2005).
\bibitem{Cramer1999} H. Cram\'{e}r, {\textit{Mathematical methods of statistics}}, vol. 9 (Princeton university press, 1999).
\bibitem{Fisher1922} R. A. Fisher, {Phil. Trans. R. Soc. A} \textbf{222}, 309 (1922); {Proc. Cambridge Philos. Soc.} \textbf{22}, 700 (1925)
\bibitem{Rao1973} C. R. Rao, {\textit{Linear Statistical Inference and Its Applications}}, (John Wiley \& Sons, New York, 1973).
\bibitem{Ruostekoski2009} J. Ruostekoski, C. J. Foot, and A. B. Deb, {Phys. Rev. Lett.} \textbf{103}, 170404 (2009).
\bibitem{Kucsko2013} G. Kucsko, P. Maurer, N. Yao, M. Kubo, H. Noh, P. Lo, H. Park, and M. Lukin, {Nature (London)} \textbf{500}, 54 (2013).
\bibitem{Toyli2013} D. M. Toyli, F. Charles, D. J. Christle, V. V. Dobrovitski, and D. D. Awschalom, {Proc. Natl. Acad. Sci. U.S.A.} \textbf{110}, 8417 (2013).
\bibitem{Jiang2003} Z. Jiang et al., {Appl. Phys. Lett.} \textbf{83}, 2190 (2003).
\bibitem{Correa2015} L. A. Correa, M. Mehboudi, G. Adesso, and A. Sanpera, {Phys. Rev. Lett.} \textbf{114}, 220405 (2015).
\bibitem{Brunelli2011} M. Brunelli, S. Olivares, M. G. A. Paris, {Phys. Rev. A} \textbf{84}, 032105 (2011).
\bibitem{Brunelli2012} M. Brunelli, S. Olivares, M. Paternostro, M. G. A. Paris, {Phys. Rev. A} \textbf{86}, 012125 (2012).
\bibitem{Jevtic2015} S. Jevtic, D. Newman, T. Rudolph, T. M. Stace, {Phys. Rev. A} \textbf{91}, 012331 (2015).
\bibitem{Higgins2013} K. D. B. Higgins, B. W. Lovett, and E. M. Gauger, {Phys. Rev. B} \textbf{88}, 155409 (2013).
\bibitem{Bruderer2006} M. Bruderer and D. Jaksch, {New J. Phys.} \textbf{8}, 87 (2006).
\bibitem{Sabin2014} C. Sab\'{i}n, A. White, L. Hackermuller, and I. Fuentes, {Sci. Rep.} \textbf{4}, 6436 (2014).
\bibitem{Stace2010} T. M. Stace, {Phys. Rev. A} \textbf{82}, 011611(R) (2010).
\bibitem{Jarzyna2014} M. Jarzyna and M. Zwierz, arXiv:1412.5609 (2014).
\bibitem{Pasquale2015} A. D. Pasquale, D. Rossini, R. Fazio and V. Giovannetti, arXiv:1504.07787 (2015).
\bibitem{Kliesch2014} M. Kliesch, C. Gogolin, M. J. Kastoryano, A. Riera, and J. Eisert, {Phys. Rev. X} \textbf{4}, 031019 (2014).
\bibitem{Boixo2007} S. Boixo, S. T. Flammia, C. M. Caves and J. M. Geremia, {Phys. Rev. Lett.} \textbf{98}, 090401 (2007).
\bibitem{Choi2008} S. Choi and B. Sundaram, {Phys. Rev. A} \textbf{77}, 053613 (2008).
\bibitem{Roy2008} S. M. Roy and S. L. Braunstein, {Phys. Rev. Lett.} \textbf{100}, 220501 (2008).
\bibitem{Mehboudi2015} M. Mehboudi, M. Moreno-Cardoner, G. De Chiara, and A. Sanpera, {New J. Phys.} \textbf{17}, 055020 (2015).
\bibitem{Gross1982} M. Gross and S. Haroche, {Phys. Rep.} \textbf{93}, 301 (1982).
\bibitem{Higgins2014} K. D. B. Higgins, S. C. Benjamin, T. M. Stace, G. J. Miburn, B. W. Lovett, E. M. Gauger, {Nat. Commun.} \textbf{5}, 4705 (2014).
\bibitem{Breuer2002} H. Breuer and F. Petruccione, {\textit{The Theory of Open Quantum Systems}}, (Oxford University Press, USA, 2002).
\bibitem{Escher2011} B. M. Escher, R. L. de Matos Filho, and L. Davidovich, {Nature Phys.} \textbf{7}, 406 (2011).
\bibitem{Demkowicz2012} R. Demkowicz-Dobrza\'{n}ski, J. Ko{\l}ody\'{n}ski, and M. Gu\c{t}\v{a}, {Nat. Commun.} \textbf{3}, 1063 (2012); J. Ko{\l}ody\'{n}ski and R. Demkowicz-Dobrza\'{n}ski, {New J. Phys.} \textbf{15}, 073043 (2013).
\bibitem{Zanardi2008} P. Zanardi, M. G. A. Paris and L. Campos Venuti, {Phys. Rev. A} \textbf{78}, 042105 (2008).
\bibitem{Haupt2014} F. Haupt, A. Imamoglu, and M. Kroner, {Phys. Rev. Applied} \textbf{2}, 024001 (2014).
\end{thebibliography}
\end{document}